\documentclass[12pt]{iopart}
\usepackage{amssymb}
\usepackage{graphicx}

\def\s0{{\sigma_0}}
\def\Tr{\mbox{Tr}~}

\newcommand{\bv}[1]{\mbox{\boldmath$#1$}}
\begin{document}

\title{Efficient entanglement operator for a multi-qubit system}


\author{Chiara Bagnasco$^1$, Yasushi Kondo$^{1,2}$ and Mikio Nakahara$^{1,2}$}
\address{$^1$ Research Center for Quantum Computing, Interdisciplinary Graduate School of Science and Engineering,
Kinki University, Higashi-Osaka, 577-8502, Japan}
\address{$^2$ Department of Physics, Kinki University, Higashi-Osaka, 577-8502, Japan}


\date{\today}

\begin{abstract}
In liquid-state NMR quantum computation,
a selective entanglement operator between qubits 2 and 3 of a three-qubit molecule is conventionally
realized by applying a pair of short $\pi$-pulses to qubit 1.
This method, called refocusing, is well suited for heteronuclear molecules. When the 
molecule is homonuclear, however, the $\pi$-pulses applied to qubit~1 
often induce unwanted 
$z$-rotations on qubits 2 and 3, even if the $z$-components of qubits 2 and 3 are left unchanged. This phenomenon is known as the 
transient Bloch-Siegert
effect, and compensation thereof is required for precise gate operation. 
We propose an alternative refocusing method, 
in which a weak square pulse is applied to qubit 1. This technique has the
advantage of curbing the Bloch-Siegert effect, making it suitable for both 
hetero- and homonuclear molecules. 
\end{abstract}

\pacs{03.67.-a, 03.65.Ud, 33.25.+k}

\maketitle

\section{Introduction}

In liquid state NMR Quantum Computing (NMR QC),
two-qubit gates are implemented through
the $J$-coupling between spins. Throughout this paper, we assume that 
the $J$-coupling tensor is isotropic in an isotropic liquid, and hence
represented by
a scalar coupling constant. To realize a selective two-qubit gate in a 
system with more than two spins, 
it is necessary to effectively suppress those spin-spin 
interactions that do not participate in gate operation.
Consider for example a molecule in which three linearly
aligned spins are employed as qubits. In NMR QC, 
a selective two-qubit gate between qubits 2 and 3 is conventionally implemented by a refocusing procedure
\cite{mn:ref:qcbook, mn:0:ref:nc, vandersypen2} in which a pair of hard
(i.e., short)
$\pi$-pulses are applied to qubit 1. This method works well for 
heteronuclear molecules. When the molecule is homonuclear, however, the 
hard pulses applied to qubit 1 often induce unwanted 
$z$-rotations on qubits 2 and 3, even if the $z$-components of qubits 2 and 3 are left unchanged. This phenomenon is 
known as the transient Bloch-Siegert (BS) effect \cite{mn:ref:qcbook, vandersypen2, bs,  
ramsey, ems, kondo, knt}. Since only a few spin one-half nuclear species 
suitable for NMR QC are known, a fully heteronuclear molecule with a large 
number of qubits is unfeasible; quantum computers with more than three qubits 
usually involve homonuclear dynamics \cite{jones}. Quantification of and 
compensation for the BS shifts are therefore essential for precise gate 
operation. 

This paper is organized as follows.
In section~\ref{sec:hardpulses}, we summarize the standard 
refocusing technique and the associated issues. 
In section~\ref{sec:proposal}, we propose an alternative method to obtain a 
selective two-qubit gate by applying a weak square pulse to qubit 1.  
We show that the BS effect is significantly reduced due to the
small ratio of the pulse amplitude and the detuning parameter,
making this method suitable for both hetero- and homonuclear molecules. 
In section~\ref{sec:fidelity}, we relax some of our assumptions to consider the full time evolution operator; we evaluate the propagator fidelity for the soft pulse method, and compare it with the fidelity obtained by numerical optimization of the conventional refocusing scheme.
In section~\ref{sec:experiment} we provide a concrete example of an experiment in which we employed the proposed soft pulse. In section~\ref{sec:conclusions} we summarize our conclusions.

%
\section{Refocusing with hard pulses}
\label{sec:hardpulses}

We consider a three-spin linear chain molecule. A radio frequency (rf)
field with a tunable amplitude $\omega_{1}$ is applied along the $x$-axis of 
qubit 1; for the time being, we will ignore the coupling between 
the rf-field and qubits 2 and 3. 
The relevant Hamiltonian of the molecule in the rotating frame of each qubit is
\begin{eqnarray}\label{eq:h}
H(\omega_{1})= \omega_{1} I_x \otimes I_2 \otimes I_2+
J_{12} I_z \otimes I_z \otimes I_2 + J_{23} I_2 \otimes I_z\otimes I_z . 
\end{eqnarray}
Here $I_2$ is the unit matrix 
of order 2, and $I_k = \sigma_k/2$, where with $\sigma_k$ ($k=x,y,z$) 
we denote the components of the Pauli vector. 

The spin-spin coupling strengths
$J_{12}$ and $J_{23}$ are fixed and always active in NMR QC.
Throughout this paper, we assume that the interaction between spins 1 and 3  ($J_{13}$) is
negligibly small.

Suppose we want to apply a two-qubit gate between spins 2 and 3. Then we need
an entanglement operator of the form
\begin{equation}
U_{23}(\alpha) = \exp(-i \alpha I_2 \otimes I_z\otimes I_z),
\label{gate}
\end{equation}
where the nonvanishing constant $\alpha$ depends on the particular gate
we are to implement (see, for example, \cite{mn:ref:qcbook}).
Free evolution ($\omega_{1}=0$) of the system under the 
Hamiltonian (\ref{eq:h}) for a duration $t^*=\alpha/J_{23}$
generates 
\begin{eqnarray}
\tilde{U}_{23}(\alpha) = e^{-i H(0) t^*} 
= e^{-i J_{12} I_z \otimes I_z \otimes I_2  t^*} U_{23}(\alpha).
\label{error}
\end{eqnarray}
To implement the operator (\ref{gate}), we need to remove the first factor 
in the right hand side of (\ref{error}) by effectively eliminating the 
action of the $J_{12}$ coupling term. A standard NMR QC refocusing approach is
to apply a pair of $\pi$-pulses of 
duration $\tau=\pi/\omega_1$ along the $x$-axis of the first spin,
separated by a time interval 
$\Delta t= (t^*/2-\tau)$ of free precession. The time evolution reads,

\begin{equation}\label{eq:ps}
U_{\rm ref}(t^*)=e^{-i H(\omega_1)\tau} e^{-i H(0) (t^*/2-\tau)}e^{-i H(\omega_1)\tau}e^{-i H(0)(t^*/2-\tau)}.
\end{equation}
If the $\pi$-pulses are `hard', {\it i.e.}, so short that the $J$-coupling time 
evolution during the application of each pulse is negligible, (\ref{eq:ps}) is reduced to
\begin{equation}\label{eq:kill}
 X^{2}e^{-i H(0) t^*/2} X^2 e^{-i H(0) t^*/2} 
= -e^{-i \alpha I_2 \otimes I_z\otimes I_z}.
\end{equation}
Here $X^2=e^{-i \pi I_x \otimes I_2 \otimes I_2}$ denotes a $\pi$-pulse 
applied along the $x$-axis of the first spin,
generated by the first term of the Hamiltonian (\ref{eq:h}).
Equation~(\ref{eq:kill}) shows that, in the vanishing pulse width limit, the unwanted factor in the right hand side of (\ref{error}) is completely 
removed. Note that the global phase factor $-1$ is irrelevant. 
This scheme works well for heteronuclear molecules, for which the Larmor 
frequencies of the spins are widely different, and hard pulses applied to
qubit 1 have practically no crosstalk to the remaining qubits. 

When the molecule is homonuclear, on the other hand, the couplings between the
rf-field and qubits 2 and 3 must be taken into account.
Then the $\pi$-pulses often induce unwanted $z$-rotations in 
qubits 2 and 3 (transient Bloch-Siegert 
effect). Suppose an $X^2$ pulse with duration $\tau$ and amplitude $\omega_{1}$ is applied to spin 1. Let $\delta_{1k}=\omega_{01}-\omega_{0k}$ ($k=2,3$) be
the difference between the Larmor frequencies of qubits 1 and $k$, and let 
$\epsilon_k = \omega_{1}/\delta_{1k}$.
We require $\tau > 1/|\delta_{1k}|$ so that the pulse is localized enough in 
the frequency domain compared to $|\delta_{1k}|$ and, at the same time,
$\tau\ll \min(1/J_{12}, 1/J_{23})$ so that the 
effect of the $J$-coupling on the time evolution is negligible for the duration of each pulse. The latter condition is typically satisfied to a
first approximation for both hetero- and homonuclear molecules (see, for example, \cite{vandersypen2,kondo, knt}).

To derive the BS phase, we describe the system in the frame rotating with
angular velocity $\omega_{01}$ - the Larmor frequency of qubit 1 -
which we call the common rotating frame \cite{knt}.
The $X^2$-pulse has the rf-frequency $\omega_{\rm rf}=\omega_{01}$.
Looked upon from qubit $k$ ($k=2,3$), whose Larmor frequency is 
$\omega_{0k}$, the rf-pulse is detuned from $\omega_{0k}$ by
$\delta_{1k}$. The effective one-qubit
Hamiltonian acting on qubit $k$ in this frame is 
\begin{equation}
\tilde{H} = \delta_{1k} I_z + \omega_1 I_x = \delta_{1k} (\epsilon_k I_x
+ I_z) = \delta_{1k} \sqrt{1+\epsilon_k^2}  \,  \hat{\bv{n}}\cdot \bv{I},
\end{equation} 
where 
$$
\hat{\bv{n}} = \frac{1}{\sqrt{1+\epsilon_k^2}}(\epsilon_k, 0, 1).
$$
Suppose the detuning $\delta_{1k}$ is large enough compared to $\omega_1$
so that $|\epsilon_k| \ll 1$. Then it follows that $\hat{\bv{n}} \simeq 
(0,0,1)$, and the time evolution operator acting on qubit $k$ 
in this frame takes the form
\begin{equation}
\tilde{U}(t) \simeq e^{-i \delta_{1k} \sqrt{1+ \epsilon_k^2} I_z t} \simeq
e^{-i \delta_{1k}I_z t} e^{-i \delta_{1k} \epsilon_k^2 I_z t/2},
\end{equation}
where we kept $\epsilon_k^2$ in the exponent since time
$t$ can be a large number.
One might naively think the detuning $\delta_{1k}$ brings about the unitary 
operator $e^{-i \delta_{1k} I_z t}$ acting on qubit $k$
in the frame rotating with $\omega_{\rm rf}
= \omega_{01}$. In reality, however, the rf-field applied to qubit 1 induces
an extra rotation angle $\delta_{1k} \epsilon_k^2 t/2$ around the $z$-axis of
qubit $k$, which affects the coordinate system fixed to qubit $k$. 
One must program the NMR spectrometer so that this 
extra angle is properly taken into account. 
Let us suppose
a $\pi$-pulse is applied to spin 1 with an amplitude $\omega_1$ and frequency 
$\omega_{\rm rf}=\omega_{01}$. The time required to implement a $\pi$-pulse is
$\tau = \pi/|\omega_1|$, from which the BS phase shift for qubit $k$ is evaluated
as $\Delta\phi_{\rm BS} = \delta_{1k} \epsilon_k^2 \pi/2|\omega_1| 
= |\omega_1|\pi/2 \delta_{1k}$. 

\begin{figure}
\begin{center}
\includegraphics[width=4cm]{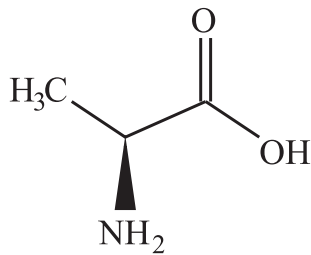}
\end{center}
\caption{
\label{fig:alanine}
Structure of L-alanine: we label the methyl carbon
as qubit 1, the $\alpha$ carbon as qubit 2, and the carboxyl carbon as qubit 3.}
\end{figure}

As a concrete example, let us evaluate the BS phase shifts 
induced by a refocusing $\pi$-pulse sequence on $^{13}$C-labeled L-alanine (figure~\ref{fig:alanine}) solved in D$_2$O. 
Three aligned carbon nuclei are employed
as qubits: the methyl carbon is labeled
as qubit 1, the $\alpha$ carbon as qubit 2, and the carboxyl carbon as qubit 3.
With these conventions, we have parameters
$J_{12}/2\pi = (34.8\pm 0.2)$~Hz, $J_{23}/2\pi = (53.8\pm 0.2)$~Hz, 
$\delta_{12}/2\pi = -4.32$~kHz, and $\delta_{13}/2\pi = -20.1$~kHz, where
the Larmor frequency of a hydrogen nucleus is $500$~MHz and 
$J_{13}$ is negligibly small \cite{kondo, kondo3}. A $\pi$-pulse with width $\tau \simeq 
0.700$~ms, which satisfies 
$1/\delta_{1k}<\tau\ll \min(1/J_{12}, 1/J_{23})$, and amplitude  
$\omega_{1} /2\pi \simeq 714$~Hz, so that $\omega_{1} \tau= \pi$,
applied to qubit 1 induces the BS phase shifts
$\pi^2/(2\delta_{12} \tau) \simeq - 0.260$~rad on qubit 2 and 
$\pi^2/(2\delta_{13} \tau) \simeq - 0.0559$~rad on qubit 3. Note that this pulse corresponds to a ``hard'' pulse in the case of a heteronuclear molecule.
Considering both pulses involved in the refocusing sequence, we find the total BS phase shifts
$\pi^2/(\delta_{12} \tau) \simeq - 0.519$~rad for qubit 2 and 
$\pi^2/(\delta_{13} \tau) \simeq - 0.112$~rad for qubit 3. 
Clearly, these BS phase shifts are sizable and must be properly 
taken into account for precise gate operation. The BS effect is usually ``compensated'' for by book-keeping of the $z$-rotations, so that the phases of the following pulses are adjusted accordingly \cite{cory}.

\section{Cancellation with Soft Pulse}
\label{sec:proposal}

We now propose an alternative implementation of the selective 
two-qubit operator $U_{23}(\alpha)$.
This method has the merit of effectively curbing the BS effect, 
making it suitable for use with homonuclear molecules.
Let us apply to qubit 1 a weak square pulse along the $x$-axis with 
duration $\tau = \alpha/J_{23}$ and a 
small amplitude $\omega_{1}$, the value of which will be fixed later
so as to eliminate unwanted time evolution.

Let us take the Hamiltonian (\ref{eq:h}) with constant $\omega_{1}
\neq 0$. The time evolution generated by this Hamiltonian
 for a time $\tau=\alpha/J_{23}$ is
\begin{equation}
e^{-i H(\omega_1)\tau} =
 e^{-i (\omega_{1} I_x \otimes I_2 \otimes I_2 + J_{12} I_z \otimes I_z \otimes I_2 )\tau}U_{23}(\alpha).
\end{equation}
We seek $\omega_1$ and $\phi$ such that
\begin{equation}\label{eq:match}
e^{-i (\omega_1
I_x \otimes I_2 \otimes I_2+ J_{12} I_z \otimes I_z \otimes I_2)\tau} = e^{i \phi} I_8
\end{equation}
is satisfied, where $e^{i \phi}$ is an irrelevant global phase. 
Since the exponent of the left hand side of (\ref{eq:match}) is traceless, 
the right hand side must be an element of SU(8) and hence
the phase
is restricted to the form $\phi = 2\pi k/8, k \in \mathbb{Z}$.
By explicitly evaluating the left hand side of
(\ref{eq:match}), we find that only 16 out of 64 matrix elements do not
vanish in general. These nontrivial equalities are reduced to the following
two equations
\begin{equation}
\sin \left(\frac{\alpha}{4} \sqrt{\frac{J_{12}^2}{J_{23}^2} 
+ \frac{4 \omega_1^2}{J_{23}^2}} \right) =0,
\ \cos \left( \frac{\alpha}{4} \sqrt{\frac{J_{12}^2}{J_{23}^2} 
+ \frac{4 \omega_1^2}{J_{23}^2}} \right) =\pm 1.
\end{equation}
The solutions are $\omega_1 = \omega_{\pm}$, where
\begin{equation}\label{soln}
\omega_{\pm} = \pm \sqrt{\frac{4 \pi^2 n^2 J_{23}^2}{\alpha^2}
- \frac{J_{12}^2}{4}},\quad n \in \mathbb{N}.
\end{equation}

To minimize the BS effect, the magnitude of which is proportional to $\omega_1$, $n$ should be the smallest integer such that the radicand of (\ref{soln}) is positive.
It turns out that $n=1$ for $J_{ij}$ of L-alanine, which we will 
consider in the following. 

Finally, by applying an rf-field $\omega_{1} = \omega_{\pm}$ for a duration $\tau = \alpha/J_{23}$ we obtain the desired operator $U_{23}(\alpha)$ up to an irrelevant global phase factor. 
Since $\omega_1=\omega_{\pm}$ is considerably smaller than the amplitude 
of the conventional hard pulses, we expect that the BS effect will be less severe.

Take $\alpha = \pi$, for example, and consider a soft pulse with width 
$\tau=\pi/J_{23} \simeq 9.29$~ms applied to qubit 1 of a deuterated 
L-alanine molecule (see section~\ref{sec:hardpulses}). We obtain 
$\omega_1/2\pi \simeq 106$~Hz and find the BS shifts 
$\omega_1^2 \tau /2 \delta_{12} \simeq - 0.0762$~rad for qubit 2 and $\omega_1^2 \tau /2 \delta_{13} \simeq -0.0164$~rad for qubit 3. Note that we do not need
to multiply these phases by 2, since there is only a single pulse applied
this time. These results are considerably smaller than those produced by a pair of hard $\pi$-pulses as shown in section~\ref{sec:hardpulses}.
In fact, these small phase shifts are comparable to experimental errors and
we may simply ignore them in designing quantum gates, which makes pulse
programming much easier than with the conventional refocusing pulses.

\section{Fidelity}
\label{sec:fidelity}

We have shown that, when we want to entangle spins 2 and 3, 
unwanted time development due to the coupling $J_{12}$ can be eliminated
by applying a soft pulse to qubit 1 rather than applying a pair of 
hard $\pi$-pulses. Note, however, that there have been certain 
oversimplifications in our analysis: for example, we have ignored the coupling between 
the rf-field and qubits 2 and 3.
We shall now lift some of these assumptions, and employ the full Hamiltonian to evaluate the propagator fidelity for the soft pulse method, and compare it with the fidelity obtained by numerically optimizing the refocusing scheme described in section~\ref{sec:hardpulses}.

Let
\begin{eqnarray}\label{eq:h123}
\hat{H}(\omega_1) &=& \omega_1 (I_x \otimes I_2 \otimes I_2+I_2
\otimes I_x \otimes I_2+I_2 \otimes I_2 \otimes I_x)  \nonumber \\ 
& & + \delta_{12} I_2 \otimes I_z \otimes I_2 + \delta_{13} 
I_2 \otimes I_2 \otimes I_z  \nonumber\\
 &&+  J_{12} I_z \otimes I_z \otimes I_2 + J_{23} I_2 \otimes I_z \otimes I_z
\end{eqnarray}
be the total Hamiltonian in the common frame rotating with the 
angular frequency $\omega_{01}$. Here we include the couplings between the 
rf-pulse with the amplitude $\omega_1$ and qubits 1, 2 and 3. 
For definiteness, let us take $\alpha = \pi$ again and say we would like to implement
an entanglement operator
\begin{equation}\label{eq:uc}
U_{23}'(\pi)= e^{-i \pi (I_2 \otimes I_z \otimes I_z)}
e^{- i \delta_{12} (\pi / J_{23}) (I_2 \otimes I_z \otimes I_2)}
e^{-i \delta_{13} (\pi / J_{23}) (I_2 \otimes I_2 \otimes I_z)}
\end{equation}
in the common rotating frame. 
This case ($\alpha =\pi$) is of special interest to us, since it produces the entangling operation for the CNOT gate.
Operator (\ref{eq:uc}) reduces to 
\begin{equation}\label{eq:u}
U_{23}(\pi) = \exp(-i \pi I_2 \otimes I_z\otimes I_z)
\end{equation}
in the individual rotating frame, in which each qubit $k$ is described in a frame rotating with the angular velocity $\omega_{0k}$.

Let us denote with $V(\tau, \omega_1)$ the propagator generated by the refocusing sequence (section~\ref{sec:hardpulses}),
\begin{equation}\label{eq:v}
V(\tau, \omega_1) =
e^{-i {\hat{H}}(\omega_1) \tau} e^{-i {\hat{H}}(0) (\pi/(2 J_{23})-\tau)}
e^{-i  {\hat{H}}(\omega_1) \tau} e^{-i {\hat{H}}(0) (\pi/(2 J_{23})-\tau)}
\end{equation}
by employing the Hamiltonian~(\ref{eq:h123}). 
Here $\tau$ and $\omega_{1}$ denote the duration and the amplitude of
each rf-pulse, respectively, and the whole process is assumed to take a time $\pi/J_{23}$ as before. In a conventional setup, 
$\omega_1$ is taken as $\pi/\tau$.
Here, however, we take $\tau$ and $\omega_1$ to be independent parameters chosen
so that they maximize the propagator fidelity defined below.
$V(\tau, \omega_1)$ consists of four processes:
(1) free evolution for a duration $\pi/(2J_{23})-\tau$;
(2) evolution under the pulse for a duration $\tau$; 
(3) free evolution for a duration $\pi/(2J_{23})-\tau$;
(4) evolution under the pulse for a duration $\tau$.
For vanishingly small values of $\tau$, with $1/|\delta_{1k}|<\tau \ll {\rm min}(1/J_{12}, 1/J_{23})$, we expect to find the conventional refocusing scheme with hard $\pi$-pulses; then, in an ideal case in which the BS effect were negligible, this propagator would produce the desired entanglement operator.

\begin{figure}
\begin{center}
\includegraphics[width=11cm]{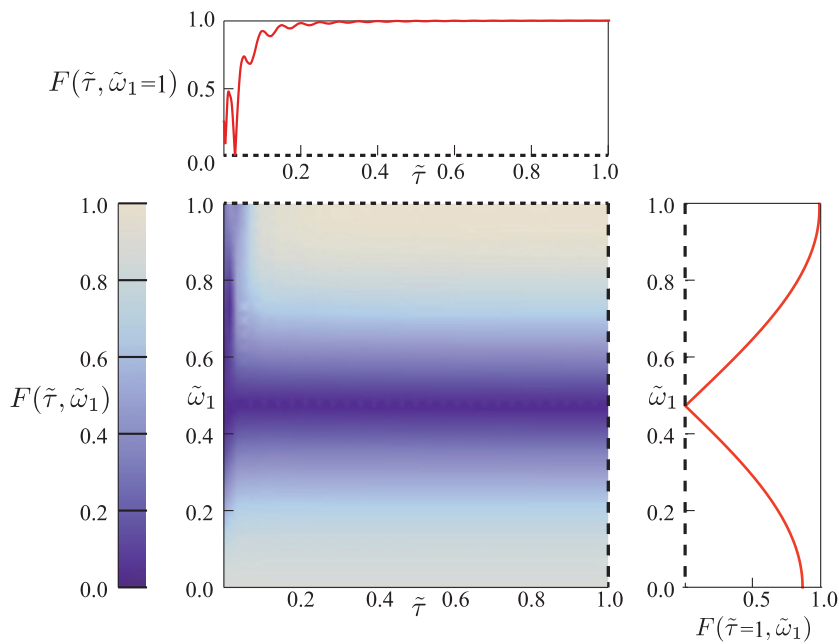} 
\end{center}
\caption{\label{fig:fidopt}(Color online)
Density plot of the propagator fidelity $F(\tilde{\tau},\tilde{\omega}_{1})$
in the domain $0 \leq \tilde{\tau} \leq 1$ and 
$0 \leq \tilde{\omega}_1 \leq 1$ (lower central panel)
for the case of deuterated $^{13}$C-labeled L-alanine. 
The left panel depicts the scale marks of the function $F\ (0 \leq F \leq 1)$ 
while the right panel
shows $F(\tilde{\tau}=1, \tilde{\omega}_1)$, the fidelity
plotted along the dashed line in the lower central panel.
The upper panel shows $F(\tilde{\tau}, \tilde{\omega}_1 =1)$,
the fidelity plotted along the dotted line in the central panel,
corresponding to the conventional refocusing scheme.
The global maximum is found
for $\tilde{\tau} \simeq 0.947$ and $\tilde{\omega}_1 \simeq 0.987$, where
$F(\tilde{\tau}, \tilde{\omega}_1) \simeq 0.999$.
}
\end{figure}

To compare the unitary matrix resulting from the process (\ref{eq:v}) with 
the target operator (\ref{eq:uc}), we define the propagator fidelity
\begin{equation}\label{eq:fid1}
F(\tilde{\tau}, \tilde{\omega}_1)=|\Tr(U_{23}^{'\dagger} (\pi)V(\tilde{\tau}, 
\tilde{\omega}_1))|/2^3,
\end{equation} 
where we have introduced dimensionless parameters $\tilde{\tau} = 2 J_{23} 
\tau/\pi$ and $\tilde{\omega}_1= \omega_1 \tau/\pi$.
We resort to numerical optimization in order to find the values of 
$\tilde{\tau}$ and $\tilde{\omega}_1$ that maximize the fidelity. 
Figure~\ref{fig:fidopt} shows the fidelity (\ref{eq:fid1})
as a function of the normalized 
pulse width $\tilde{\tau}$ ($0 \leq \tilde{\tau} \leq 1$)
and the normalized amplitude $\tilde{\omega}_{1}$
($0 \leq \tilde{\omega}_1 \leq 1$)
for the case of deuterated $^{13}$C-labeled L-alanine (see section~\ref{sec:hardpulses}). We calculate 
that the global optimal result is given by $F_{\rm opt}\simeq 0.999$ for $\tilde{\tau}
\simeq 0.947$ ($\tau \simeq 4.40$~ms)
and $\tilde{\omega}_1 \simeq 0.987$ ($\omega_1/2\pi  \simeq 112$~Hz). 

The conventional refocusing scheme is retrieved by setting $\tilde{\omega}_1
=1$ (upper central panel in figure~\ref{fig:fidopt}). In this case, small values
of $\tau$ in the interval $1/|\delta_{1k}|< 
\tau \ll {\rm min}(1/J_{12}, 1/J_{23})$ correspond to the conventional refocusing scheme with a pair of hard pulses; in particular, for the case of two ``hard'' $\pi$-pulses with $\tilde{\tau} \simeq 
0.151$ ($\tau=0.700$~ms (section~\ref{sec:hardpulses})), we find $F(0.151,1)\simeq 0.965$. 
As $\tilde{\tau}$ approaches 1 (so that $\tau$ approaches $\pi/2J_{23}$), the fidelity oscillates 
slightly about the value $F(1,1)\simeq 0.998$. Let us note that for $\tilde{\omega}_1
=1$ and $\tilde{\tau}=1$ ($\tau=\pi/2J_{23}\simeq 4.65$~ms,
$\omega_{1} /2\pi \simeq 108$~Hz), the two pulses are merged together to form a single
$2\pi$-pulse: this choice corresponds to the soft pulse case with a slightly detuned $\omega_1$ (see below).

The fidelity for the soft pulse solution obtained in section~\ref{sec:proposal} is
easily evaluated by setting 
$\omega_{1} /2\pi=\omega_{\rm1sp}/2\pi
\simeq 106$~Hz 
(with $\omega_{\rm1sp}= \sqrt{4 J_{23}^2 -{J_{12}^2}/{4}}$, $\tilde{\omega}_1=\tilde{\omega}_{\rm1sp} \simeq 0.987$)
in the Hamiltonian (\ref{eq:h123}) and $\tilde{\tau} =1$,
resulting in $F(1,\tilde{\omega}_{\rm1sp}) \simeq 0.999$.

We find that, 
according to our simulations, the fidelity for the soft pulse scheme (0.999) is better than that obtained with the standard refocusing scheme (0.965) employing hard pulses, and comparable with the fidelity obtained by numerical optimization; moreover, the parameters $\omega_1$ and $\tau$ for the soft pulse are conveniently derived from the knowledge of $J_{12}$ and $J_{23}$.

\section{Experimental implementation}
\label{sec:experiment}

\begin{figure}
\begin{center}
\includegraphics[width=8cm]{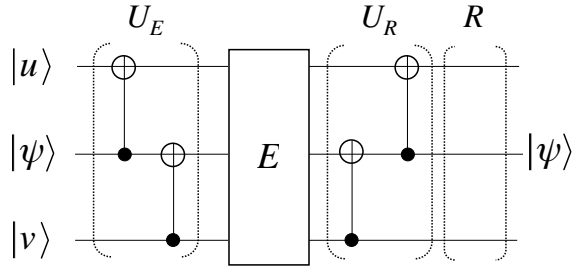}
\end{center}
\caption{\label{fig:qcnmr}
Quantum circuit used in experimental implementation with three-qubit NMR Quantum Computer. The information to be protected is carried by the second qubit.}
\end{figure}

Let us now provide a concrete example in which we made practical use of the soft pulse technique described above. Consider a system of three qubits, which are all simultaneously afftected by an external noise represented by the {\it fully correlated} error channel 

\begin{eqnarray}
{\mathcal E} (\rho) 
&=& \sum_{i=0}^3 p_i E_i (\rho) E_i^\dagger,
\label{eq:errorch}
\end{eqnarray}
where $E_0 = \sigma_0^{\otimes 3}, E_1 = \sigma_x^{\otimes 3}, 
E_2 = \sigma_y^{\otimes 3}, E_3 = \sigma_z^{\otimes 3}$. 
The operators $\{ E_k \}$ are the Kraus operators (or errors) associated with ${\mathcal E}$.  Here $p_i \ge 0 $ is the probability with which an error operator $E_i$
acts on the quantum system with density matrix $\rho$, and we assume $\sum_{i=0}^3 p_i =1$.
In our recent work \cite{qec}, we proposed a simple operator quantum error correction scheme which protects one data qubit against this type of noise by encoding it with two ancilla qubits in an arbitrary mixed state. The encoding operator $\mathcal{U}_E$ and the decoding operator 
$\mathcal{U}_R$ 
are implemented with two CNOT gates each. We proved \cite{qec} that this scheme provides the simplest noiseless subsystem, in terms of the number of CNOT gates, under our noise model.
We implemented this scheme experimentally using a three-qubit NMR quantum computer, in which the ancillae are in the maximally mixed state. We employed a JEOL ECA-500 NMR spectrometer, whose hydrogen Larmor frequency is approximately 500~MHz. 
As a linear chain molecule with three coupled spins to be used as qubits, we employed ${}^{13}$C-labeled L-alanine (98\% purity, Cambridge Isotope) solved in D$_2$O.
The quantum circuit takes the form shown in figure~\ref{fig:qcnmr}, wherein we designated the second qubit as the data qubit carrying the information to be protected. 

If we denote the ancillae as $|u \rangle$, $|v\rangle$, and the data qubit as $|\psi \rangle$, it can be shown that,
\begin{eqnarray}
\Tr_{1,3}(\mathcal{U}_R \circ {\mathcal E} \circ \mathcal{U}_E )
( |u\rangle \langle u| \otimes |\psi \rangle \langle\psi| \otimes |v \rangle\langle v|)=
|\psi \rangle \langle\psi|,
\label{eq:CB-ns_s}
\end{eqnarray}
for any $|u \rangle$, $|v\rangle$, where $\Tr_{1,3}$ denotes the partial trace over qubits 1 and 3.

In the experimental pulse sequences realizing the encoding and decoding operations, we  employed soft pulses to implement the two two-qubit gates for each operation.

\begin{figure}
\begin{center}
\includegraphics[width=9cm]{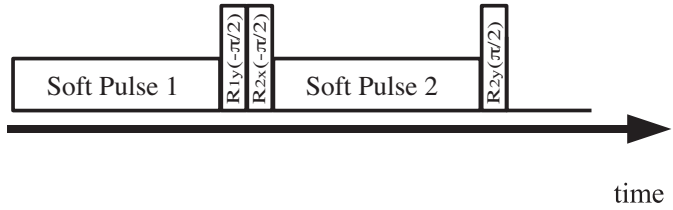}
\end{center}
\caption{\label{fig:pse}
Pulse sequence implementing the encoding operation employing soft pulses. Here $R_{\rm 1{\it y}}(- \pi/2)=e^{i (\pi/2)I_y \otimes I_2 \otimes I_2}$, for example.}
\end{figure}

We denote with $U^{\rm sp}_{ij}(\pi)$ the soft pulse operator implementing the two-qubit gate between qubits $i$ and $j$; if we neglect some irrelevant phases, the encoding operation reads (see figure~\ref{fig:pse})

\begin{equation}
\mathcal{U}^{\rm NMR}_E = 
e^{-i (\pi/2)I_2 \otimes I_y \otimes I_2} U^{\rm sp}_{23}(\pi) e^{i (\pi/2)I_2 \otimes I_x \otimes I_2} e^{i (\pi/2)I_y \otimes I_2 \otimes I_2} U^{\rm sp}_{12}(\pi),
\end{equation}
and the deconding operation is
\begin{equation}
\mathcal{U}^{\rm NMR}_R = U^{\rm sp}_{12}(\pi) e^{-i (\pi/2)I_y \otimes I_2 \otimes I_2} e^{i (\pi/2)I_2 \otimes I_x \otimes I_2} U^{\rm sp}_{23}(\pi) e^{i (\pi/2)I_2 \otimes I_y \otimes I_2}. 
\end{equation}

We find that 
\begin{eqnarray*}
\begin{array}{ccc}
 \mathcal{U}_R  E_0 \mathcal{U}_E &=&
- 4(I_z \otimes I_2 \otimes I_z),\\
 \mathcal{U}_R  E_1 \mathcal{U}_E &=&
  2i (I_2 \otimes I_2 \otimes I_x),\\
 \mathcal{U}_R  E_2 \mathcal{U}_E &=&
- 4(I_y \otimes I_2 \otimes I_x),\\
 \mathcal{U}_R  E_3 \mathcal{U}_E &=&
-4 i (I_x \otimes I_2 \otimes I_z),
\end{array}
\end{eqnarray*}
{\it i.e.}, upon retrieval, the second qubit is found not to be affected by the noise operators.

Experimental results \cite{qec} also show that the algorithm effectively protects the data qubit from the effect of fully correlated noise.

\section{Conclusions}
\label{sec:conclusions}

We consider a linear chain molecule with three coupled spins and suppose 
we want
to implement an entanglement operator
(\ref{gate})
to realize a selective two-qubit gate between spins 2 and 3.
In conventional NMR QC, this is achieved by applying a pair of hard $\pi$-pulses to qubit 1. When the molecule is homonuclear, however, one needs to take into account the Bloch-Siegert effect in designing quantum gates. We proposed an alternative method to obtain the entanglement operator (\ref{gate}) by applying a weak pulse to spin 1. Unwanted factors are removed by an appropriate choice of the rf-field amplitude $\omega_1$ and duration $\tau$. 
The BS effect for such a weak pulse is negligible in general, which makes NMR
pulse programming and quantum gates design much simpler than with
conventional hard $\pi$-pulses; it also makes this method suitable for use with
both homo- and heteronuclear molecules. We employed the proposed scheme in an operator quantum error correction experiment \cite{qec}. This technique should be also applicable to any physical system, for which the coupling constants are not controllable.

\ack

We are grateful to the `Open Research Center' Project for Private Universities, matching fund subsidy from the MEXT (Ministry of Education, Culture, Sports, Science and Technology) for financial support. 
YK and MN would like to thank
partial supports of Grants-in-Aid for Scientific Research
from the JSPS (Grant Nos. 23540470 and 25400422).
CB acknowledges financial support by the MEXT Scholarship for foreign students. 


\end{document}